\documentclass[runningheads]{llncs}
\usepackage{enumitem}
\usepackage[T1]{fontenc}
\usepackage{graphicx}
\usepackage{hyperref}

\usepackage{amssymb}

\usepackage{array}
\usepackage{mdwmath}
\usepackage{mdwtab}
\usepackage{eqparbox}
\usepackage{url}
\usepackage{listings}
\usepackage{graphicx}
\usepackage{program}
\usepackage{ifthen}
\usepackage{epsfig}
\usepackage{array}
\usepackage{color}
\usepackage[margin=1in]{geometry}
\usepackage{hyperref}
\usepackage{url}
\usepackage{xspace}
\usepackage{wasysym}
\usepackage{booktabs}
\usepackage{algorithmic}
\usepackage{textcomp}
\usepackage{comment}
\usepackage{balance}
\usepackage{subcaption}
\usepackage{multirow}

\definecolor{blue}{rgb}{0,0,1.0}
\definecolor{darkgreen}{rgb}{0,0.44,0}
\definecolor{green}{rgb}{0,0.44,0}
\definecolor{darkred}{rgb}{0.44,0,0}
\definecolor{darkblue}{rgb}{0,0,0.64}
\definecolor{mygray}{rgb}{0.9,0.9,0.9}
\definecolor{mymauve}{rgb}{0.58,0,0.82}
\definecolor{myred}{rgb}{0.72,0.18,0.0} 
\definecolor{mygreen}{rgb}    {0.0,0.72,0.0} 
\definecolor{myblue}{rgb} {0.18,0.0,0.72} 
\definecolor{mycreme}{rgb}        {1.0,0.8,0.2} 
\definecolor{mygray}{rgb}{0.95,0.95,0.95}
\definecolor{darkgray}{rgb}{0.55,0.55,0.55}

\newcommand{\gemm}{{\sc gemm}\xspace}

\begin{document}
\title{Enabling RISC-V Vector Code Generation in MLIR through Custom xDSL Lowerings}
\titlerunning{RISC-V Vector Code Generation in MLIR through Custom xDSL Lowerings}
%
\author{Jie Lei\inst{1}
\and
 H\'ector Mart\'inez\inst{2}
 \and
Adri\'an~Castell\'o\inst{1}
}
\authorrunning{J. Lei et al.}
%
\institute{Universitat Polit\`ecnica de Val\`encia, Val\`encia, Spain
 \and
Universidad de C\'ordoba, C\'ordoba, Spain
\\
\email{jlei@disca.upv.es, el2mapeh@uco.es, adcastel@disca.upv.es}
}
\maketitle              
\begin{abstract}
The growing adoption of RISC-V in high-performance and scientific computing has increased the need for performance-portable code targeting the RISC-V Vector (RVV) extension. However, current compiler infrastructures provide limited end-to-end support for generating optimized RVV code from high-level representations to low-level implementations. In particular, 
existing MLIR distributions lack practical lowering paths that map high-level abstractions to RVV intrinsics, limiting their applicability for production-ready RISC-V kernels.
This paper presents a compilation approach that combines MLIR with xDSL to bridge the missing lowering stages required for RVV code generation. Using custom intermediate representations and transformation passes implemented in xDSL, we systematically translate high-level operations into specialized, hardware-aware C code invoking RVV intrinsics. The resulting kernels are emitted as portable C functions that can be directly integrated into existing applications, enabling incremental adoption without modifying surrounding software stacks.

We demonstrate the approach on the General Matrix Multiplication (\gemm) kernel and evaluate the generated micro-kernels on two real RISC-V platforms, the K230 and the BananaPi F3, comparing against OpenBLAS for both square-matrix benchmarks and transformer-based workloads derived from the BERT-Large model. 

When integrated into a matrix multiplication kernel, the proposed approach consistently outperforms OpenBLAS, reaching up to 12.2 GFLOPS compared to the baseline’s 5.1 GFLOPS and providing performance improvements between 10–35\% across the evaluated workloads.
These results demonstrate that combining MLIR with xDSL provides a practical pathway to portable, optimized code generation for RISC-V platforms. 

\end{abstract}

\section{Introduction}
\label{sec:introduction}
The growing adoption of RISC-V as a common Instruction Set Architecture (ISA) across domains ranging from Internet-of-Things (IoT) devices to high-performance computing (HPC) systems has introduced new challenges for software portability and optimization. In the HPC context, modern RISC-V processors increasingly implement the RISC-V Vector (RVV). 
extension in diverse configurations, including Simple-Instruction Multiple-Data (SIMD) floating-point units of varying widths, and long vector register architectures. Although these platforms share a common ISA, their microarchitectural diversity complicates the development of portable, high-performance software.

Achieving high performance on RVV requires explicit management of vector-length–agnostic programming, intrinsic selection, and tile sizing. As the number and heterogeneity of RISC-V implementations continue to grow, maintaining hand-tuned kernels becomes increasingly impractical, even for regular HPC workloads such as dense linear algebra.

Compiler infrastructures such as Halide~\cite{Halide}, MLIR~\cite{mlir2,MLIR}, Apache TVM~\cite{TVM_1}, or Exo~\cite{yukaexo} address this challenge by generating optimized, hardware-aware code from high-level representations. Among them, MLIR provides a powerful multi-level intermediate representation (IR) and extensible lowering framework. However, current MLIR distributions lack complete end-to-end lowering paths for RVV targets, particularly for mapping high-level vector abstractions to RVV intrinsics and hardware-aware C code. This limitation restricts MLIR’s applicability to production-ready RVV kernel generation.

In this work, we bridge this gap by combining MLIR with xDSL~\cite{xdsl,xdsl2}, a Python-native toolkit for designing domain-specific compilers, to enable systematic generation of C code targeting RVV intrinsics. Using xDSL, we implement the missing lowering stages and introduce IRs that explicitly model RVV semantics and vector-length–aware transformations. The resulting kernels are emitted as standalone C functions that can be directly integrated into existing applications. 

Taking advantage of the automatic code generation, we also benefit from generating a complete collection of different micro-kernels. The generated combinations are then used for fine-tuning HPC applications, making it possible to choose the best micro-kernel for each scenario, and in particular for deep learning (DL) models, the best micro-kernel for each layer and target architecture.  

Concretely, this paper makes the following contributions:
\begin{itemize}[topsep=0pt]

    \item We analyze the limitations of MLIR for RVV code generation and identify the missing lowering stages.
    \item We design and implement a hybrid MLIR–xDSL compilation pipeline that bridges these gaps through custom dialects and transformations.
    \item We enable the generation of specialized, hardware-aware C micro-kernels for the General Matrix Multiplication (\gemm), targeting RVV intrinsics. 
    \item We evaluate the generated C code and compare it against the highly optimized implementation of \gemm in OpenBLAS~\cite{OpenBLAS} 
    using different DL models.
    \end{itemize}

The rest of the paper is organized as follows. Section~\ref{sec:back} summarizes a high-performance algorithm for \gemm and introduces the MLIR and xDSL frameworks; Section~\ref{sec:related} lists other approaches to automatic code generation and visits some existing work; Section~\ref{sec:generator} presents the complete workflow of our proposed solution; Section~\ref{sec:experiments} evaluates and compares the auto-generated C code on \gemm operations; Section~\ref{sec:conclusions} closes the paper with some concluding remarks.

\section{Background}
\label{sec:back}

\subsection{\gemm in High-Performance Libraries}

Consider the \gemm 
$C \leftarrow C + AB$,
where $A \in \mathbb{R}^{m \times k}$, 
$B \in \mathbb{R}^{k \times n}$, and 
$C \in \mathbb{R}^{m \times n}$.
High-performance implementations of \gemm in modern BLAS libraries
follow the algorithmic structure popularized by GotoBLAS~\cite{Goto:2008:AHP}, with the computation organized as a hierarchy of loops and packing
routines designed to exploit the memory hierarchy and SIMD units, as depicted in Figure~\ref{lst:bliscode}.

\begin{figure}[tb]
\centering
\begin{minipage}[t]{0.8\textwidth}
\begin{lstlisting}[language=C, basicstyle=\scriptsize\ttfamily, numbers=left, frame=lines]
for (jc=0; jc<n; jc+=nc)                // Loop L1
  for (pc=0; pc<k; pc+=kc) {            // L2
    Bc := B(pc:pc+kc-1,jc:jc+nc-1);     // Pack B
    for (ic=0; ic<m; ic+=mc) {          // L3
      Ac := A(ic:ic+mc-1,pc:pc+kc-1);   // Pack A
      for (jr=0; jr<nc; jr+=nr)         // L4
        for (ir=0; ir<mc; ir+=mr)       // L5
          // Micro-kernel               
          for (pr=0; pr<kc; pr++)       // L6
              C(ic+ir:ic+ir+mr-1, jc+jr:jc+jr+nr-1) 
              += Ac(ir:ir+mr-1,pr) *  Bc(pr,jr:jr+nr-1);
  }}
\end{lstlisting}
\end{minipage}
\caption{Pseudo-code of the Goto-based BLIS \gemm algorithm.}
\label{lst:bliscode}
\end{figure}

In particular, at the outmost level, three nested loops iterate over the $n$-, $k$-, and
$m$-dimensions using cache-aware blocking parameters $n_c$, $k_c$,
and $m_c$. These loops define the \emph{macro-kernel}, whose role is to 
partition the matrices into panels that fit different cache levels.
Submatrices of $A$ and $B$ are reorganized and packed into contiguous buffers
($A_c \in \mathbb{R}^{m_c \times k_c}$ and $B_c \in \mathbb{R}^{k_c \times n_c}$) to ensure predictable
cache behavior.

The actual and critical arithmetic is performed inside the \emph{micro-kernel},
a component that is typically written in assembly or in C using vector intrinsics.
The micro-kernel updates a small micro-tile of matrix $C$ of size
$m_r \times n_r$ processing the problem across the $k_c$ dimension, effectively performing
a sequence of outer-product operations.
The packed micro-panels $A_r \in \mathbb{R}^{m_r \times k_c}$
and $B_r \in \mathbb{R}^{k_c \times n_r}$, which are part of $A_c$ and $B_c$, are arranged to maximize
register reuse, ensure unit-stride memory access, and improve SIMD efficiency.

Adapting this structure to RISC-V with RVV requires
careful handling of vector-length--agnostic execution,
intrinsic selection, and tile sizing, which motivates
systematic code generation approaches.

\subsection{MLIR}

MLIR is a compiler infrastructure developed within the LLVM ecosystem to address the growing complexity of modern software stacks, particularly in domains such as HPC and machine learning. Unlike traditional compiler IRs that operate at a single abstraction level, MLIR supports multiple levels of abstraction within a unified framework. It achieves this through a flexible system of dialects, each of which can model domain-specific operations, types, and transformations. This design enables progressive lowering from high-level algorithmic representations 
to low-level hardware-oriented code while preserving semantic structure throughout the compilation pipeline.

A key strength of MLIR lies in its extensibility and composability. Developers can define custom dialects, introduce new transformation passes, and interoperate with existing LLVM backends. MLIR also provides infrastructure for pattern rewriting, canonicalization and analysis, facilitating systematic optimization and code generation. In the context of machine learning and linear algebra, MLIR enables structured representations of tensor operations, memory references, and arithmetic expressions, which can be incrementally lowered toward target-specific dialects or external code emission frameworks.

\subsection{xDSL}
xDSL is a lightweight, Python-native compiler toolkit designed to mirror and complement the MLIR philosophy while emphasizing rapid prototyping and accessibility. It provides a lightweight environment for defining dialects, constructing IRs, and implementing lowering passes without the engineering overhead of large C++-based infrastructures. By adopting many of MLIR’s core abstractions, such as operations, regions, blocks, and dialects xDSL maintains conceptual compatibility while significantly reducing development complexity.

A primary advantage of xDSL is its flexibility in experimentation and automation. Its Python foundation enables seamless integration with scripting, testing frameworks, and external tooling, making it particularly suitable for research-driven compiler development. With xDSL, developers can quickly prototype new transformations, generate IR, and iterate on lowering strategies with minimal boilerplate. As a result, xDSL serves as an effective environment for exploring novel compilation pipelines and bridging gaps in existing infrastructures. 

\section{Related Work} 
\label{sec:related}

 The rapid proliferation of computing platforms has created a significant performance portability challenge: achieving efficient execution across architectures with fundamentally different vector widths, memory hierarchies, instruction sets, and architectural characteristics. 
In this context, automatic generation of optimized code has become a central research topic. Therefore, in this section, we briefly revisit some examples applied to the generation of \gemm and micro-kernel. The authors of~\cite{alaejos2022micro,siwinska2025enhancing} use TVM to optimize the overall \gemm operation across different hardware architectures, and the authors from~\cite{GEMM_MLIR} use MLIR to describe early experiences also with the entire \gemm algorithm.

In the case of optimized micro-kernels, the work in~\cite{10444883} focuses on generating Arm NEON solutions. Moreover, the work in~\cite{castelloISCcots} uses the Exo framework to implement specialized code generators for RISC-V CPUs.  

From the perspective of using xDSL for code generation, our work is inspired by~\cite{10.1145/3696443.3708952}, which also utilizes xDSL and MLIR for RISC-V extension. Their approach explicitly avoids general-purpose compilers, opting instead to build a self-contained backend optimized for their in-house micro-architecture, namely the Snitch accelerator, to generate highly efficient native assembly code. 

In contrast, our work prioritizes broad portability and agile development across the fragmented RISC-V ecosystem, specifically targeting the RVV intrinsics. By emitting standard C code, our dynamically sized micro-kernels act as a ``universal assembly.'' This ensures wider applicability and seamless integration into existing toolchains without locking the user into a single custom hardware backend.
\section{Building the Code Generator}
\label{sec:generator}

This work aims to leverage MLIR to facilitate automatic code generation for high-performance \gemm designs on RISC-V CPUs. Therefore, the ultimate goal is to propose a design flow that allows users to specify the micro-kernel dimensions $mr \times nr$, the data type, and the length of the vector registers using high-level MLIR dialects. 

The workflow is designed to generate portable C code micro-kernels enriched with native RVV 1.0 intrinsics. Standard C code was strategically chosen as the final output format, providing a highly compatible solution. Due to its portability, it can be compiled and executed across diverse RISC-V platforms. 

\subsection{The incomplete MLIR lowering algorithm}

MLIR facilitates the C code generation through its \texttt{emitc} dialect,  created to bridge the gap between LLVM abstractions and novel or embedded systems where access to full LLVM infrastructure is limited. Once the code is fully written in \texttt{emitc}, the built-in \texttt{mlir-translate} tool directly converts it into C code.

In MLIR, users can formulate their designs using higher-level abstractions in a variety of dialects. For example, they can express floating-point multiplication 
 as \texttt{\%a = arith.mulf \%b, \%c : f32} 
using the \texttt{arith} dialect; or dynamic memory allocation 
as \texttt{\%0 = memref.alloc() : memref<?xf32, 1>} 
using the \texttt{memref} dialect. Both of these operations are critical for high-performance \gemm.

Therefore, we started using MLIR (LLVM 22.0) to lower these high-level functions to \texttt{emitc}. While this version successfully lowered the \texttt{arith} dialect, it was unable to handle the \texttt{memref} lowering. This problem occurs because the lowering pass from \texttt{memref} to \texttt{emitc} for dynamic array types is incomplete in the upstream version, preventing the generation of standard C pointers. We also tested that MLIR was able to lower code written in pure \texttt{emitc} to C. However, writing \texttt{emitc} is close to doing so in assembly, which this work aims to avoid. 

Furthermore, since the target hardware involves RISC-V systems with Vector Extensions, we have also expanded our design to integrate native RVV intrinsic support directly into the emission flow.

\subsection{The proposed MLIR–xDSL hybrid automatic code generation pipeline}

While attempting to implement our custom lowering passes directly within the upstream MLIR project, we encountered  the high development cost and rigidity of the MLIR infrastructure, which presents a steep learning curve. Focusing on our core objective, which aims to develop a lowering process that converts high-level MLIR dialects into \texttt{emitc} for generating C code via the \texttt{mlir-translate} tool,  we concluded that we needed an alternative framework to simplify the design of the lowering.

\begin{figure}[tb]
    \centering
    \includegraphics[width=0.8\linewidth]{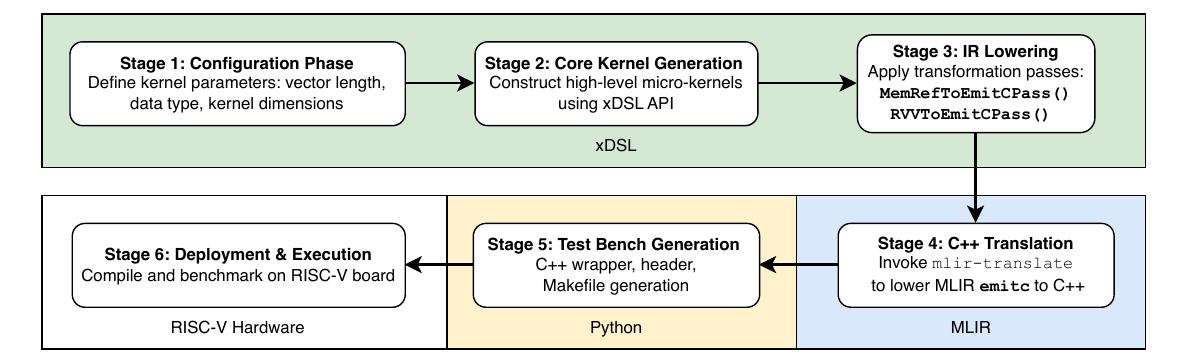}
    \caption{MLIR-xDSL hybrid code generation pipeline }
    \label{figures:xdsl_flow.pdf}
\end{figure}

To resolve this, we mixed xDSL and MLIR in the hybrid pipeline in Figure~\ref{figures:xdsl_flow.pdf}. We utilize xDSL to generate micro-kernels with a user-specified configuration in MLIR dialects. The xDSL uses the xDSL API to generate an IR design mixed with various MLIR dialects. Once the IR has been generated, it passes through our lowering algorithm to generate the \texttt{emitc} code. Then, we return to the official \texttt{mlir-translate} tool to complete the C translation workflow.

Within this unified environment, we leverage Python to automatically generate the necessary C project headers and main test programs, deploy the source files to a RISC-V board, compile them natively, and benchmark each kernel's performance. This seamless integration ensures that the end-to-end development cycle from micro-kernel high-level specification to code generation and benchmarking can be executed in a matter of minutes.

\subsection{The end-to-end micro-kernel generator pipeline}

To illustrate the capabilities of our MLIR-xDSL hybrid code generation pipeline, 
we trace the creation of an $8 \times 4$ RISC-V Vector (RVV) micro-kernel. We focus specifically on the most critical component: Loop L6 of the micro-kernel (Figure~\ref{lst:bliscode}), which performs the Fused Multiply-Accumulate (FMA) operations.

~\\ \noindent
\textbf{Stage 1: Micro-kernel configuration (xDSL).}
The pipeline is designed to be flexible for constructing kernels for both scalar and vector data types. The pipeline begins in the Python-based xDSL environment where the user specifies the micro-kernel dimensions (e.g., $m_r=8, n_r=4$), the data type (e.g., FP32), and the target vector length expressed in bits (e.g., \texttt{vlen=256}). Our custom xDSL abstractions then dynamically construct the high-level MLIR representation for all micro-kernels 
for
all possible combinations 
of dimensions
from $1\times1$ to $m_r\times n_r$. This automatically generated set of micro-kernels cover all possible edge cases in the \gemm operation. 

~\\ \noindent
\textbf{Stage 2: Core kernel generation (xDSL).}
Our Python \texttt{IRBuilder} orchestrates the construction of the micro-kernel, including loading the micro-tile from matrix $C$ to $C_r$ registers, loop L6, and storing the results from $C_r$ registers into $C$ matrix. L6 loop includes loading data from the buffer $A_c$ to $A_r$ registers and $n_r$ \texttt{AXPI} operations. Notice that the micro-kernel loads the values from the buffer $B_c$ directly instead of using $B_r$ registers due to the performance loss pointed in~\cite{castelloISCcots}.   The code for the generation of loop $L6$ is shown in Figure~\ref{lst:python_ir_builder}.

\begin{figure}[t]
\centering
\begin{minipage}[t]{\textwidth}
\begin{lstlisting}[language=python, basicstyle=\scriptsize\ttfamily, numbers=left, frame=lines]
# 1. Emit vle32.v instruction to load the Ar vector from Ac data
Ar.append(lb.vle32(Ac, k_mul_lda, vl))

# 2. Emit nr vfmacc instructions to multiply scalar from Bc across vector Ar, accumulating into Cr
for i in range(nr):
    c_new[(i, mr)] = lb.vfmacc(
        Cr[(i, mr)],   # Incoming SSA accumulator
        Bc, b_offset,  # Bc memory scalar
        Ar[mr],        # Loaded A vector
        vl             # Vector length
    )

# 3. Yield the newly minted vector registers to the next iteration
yield_vals = [c_new[(i, j)] for n in range(nr) for j in range(mr)]
loop_body.add_op(scf.YieldOp(*yield_vals))

# 4. Construct the surrounding scf.for loop representation of L6
for_op = scf.ForOp(start, stop, step, iter_inits, Region([loop_body]))
\end{lstlisting}
\caption{Python xDSL representation of loop L6 for the $8 \times 4$ micro-kernel.}
\label{lst:python_ir_builder}
\end{minipage}
\end{figure}

Line 2 emits the vector load of the $A_c$ values into $A_r$ registers. In the case of $8 \times 4$ micro-kernel, only one register is used because the 8 FP32 elements fit inside a unique register of 256~bits; lines 5--11 insert $n_r$ (4 in this case) FMA instructions using the \texttt{lb.vfmacc} abstraction that is part of our proposed \texttt{rvv} dialect; lines 14--15 add the logic for reusing the $C_r$ register accumulators (\texttt{scf.YieldOp}); and line 18 constructs the $L6$ for loop.

\noindent The code in Figure~\ref{lst:python_ir_builder} produces the MLIR IR in Figure~\ref{lst:mlir_fma_loop},  where: lines 5--6 display the first  RVV FMA operation of loop L6. We omit the other 3 FMAs' instructions for brevity; line 9 manages the reuse of the $C_r$ registers; and lines 1--10 are the body of the loop L6  (\texttt{scf.for}).
The variable naming method used in MLIR code uses sequential names like \texttt{\%33} or \texttt{v68} because the MLIR tool deterministically maps the underlying MLIR register numbering.

\begin{figure}[tb]
\centering
\begin{minipage}[t]{\textwidth}
\begin{lstlisting}[language=C, basicstyle=\scriptsize\ttfamily, numbers=left, frame=lines]
%37, ..., %40 = scf.for %41 = %10 to %0 step %11 iter_args(%42 = %19,...) -> (...) 
{
  ...
  // Emit FMA for Column 0 (taking incoming accumulator %42)
  %61 = "rvv.vfmacc_vf_f32m1Op"(%42, %4, %53, %52, %51) : 
    (!rvv.vfloat32m1, memref<-1xf32>, index, !rvv.vfloat32m1, index) -> !rvv.vfloat32m1
  ...
  // Explicit SSA handoff of new accumulators (%61) to the next loop iteration
  scf.yield %61, %62, %63, %64
}
\end{lstlisting}
\caption{MLIR IR with the \texttt{scf.for} representation of loop L6 and FMA bindings.}
\label{lst:mlir_fma_loop}
\end{minipage}
\end{figure}

To understand how the xDSL pipeline orchestrates the FMA generation, let us break down the 5 arguments passed into the generated MLIR \\\texttt{rvv.vfmacc\_vf\_f32m1Op} instruction: \texttt{\%42 (\!rvv.vfloat32m1)} is the running accumulator register $C_r$ for this column of $C$, mandated by the SSA \texttt{iter\_args}; \texttt{\%4 (memref<\dots>)}: the memory pointer to the $B_c$ packing; \texttt{\%53 (index)} is the integer offset required to extract the scalar from $B_c$; \texttt{\%52 (\!rvv.vfloat32m1)} is the previously fetched $A_r$ vector register loaded from $A_c$; and \texttt{\%51 (index)} is the vector length controlling the operation's hardware bounds.

The \texttt{rvv.*} namespace including both the FMA instruction and the \texttt{!rvv.vfloat32m1} data type is our proposed, custom RVV-MLIR dialect.

~\\ \noindent
\textbf{Stage 3: IR lowering to \texttt{emitc} (xDSL).}
The current IR structure mixing \texttt{scf}, \texttt{memref}, \texttt{arith}, and custom \texttt{rvv} dialects cannot be directly compiled. Therefore, this stage in the pipeline applies targeted transformation passes one-by-one to lower these high-level dialects into \texttt{emitc}. Concretely, as the result of this stage, our xDSL-generated MLIR IR converges into an MLIR pure \texttt{emitc} dialect code. Reaching that result is done as follows:  
First, xDSL addresses memory abstractions through the \texttt{MemRefToEmitCPass}.  
The \texttt{memref} type provides a mechanism for representing multidimensional arrays with complex strides and layouts. However, C compilers natively operate on flat memory blocks managed by standard pointers. This pass identifies \texttt{memref} occurrences (such as structures $A_c,B_c,C$) and translates them to standard C pointer abstractions. We can trace this specific transformation through the kernel function's signature across the pipeline stages summarized in Figure~\ref{lst:memref_lowering}. Line 2 defines the xDSL function with the \texttt{memref} objects; lines 4--5 show \texttt{emitc.ptr} structures; and line 9 the C pointers.

\begin{figure}[tb]
\centering
\begin{minipage}[t]{\textwidth}
\begin{lstlisting}[language=C, basicstyle=\scriptsize\ttfamily, numbers=left, frame=lines]
// 1. Initial xDSL/MLIR Generation (Abstract Vector Algebra)
func.func @xdsl_api_kernel_(%0: index, %1: memref<-1xf32>, %2: memref<-1xf32>, ...) { ... }

// 2. Transformed EmitC IR (Flattened Memory Pointers)
"func.func"() <{sym_name = "xdsl_api_kernel_", 
                function_type = (index, !emitc.ptr<f32>, !emitc.ptr<f32>, ...) -> ()}>

// 3. Final Executable C++ Emission (Standard C Arrays)
void xdsl_api_kernel_(void*, int v1, float* v2, float* v3, ...) { ... }
\end{lstlisting}
\caption{Transformation of the kernel function signature across generating stages.}
\label{lst:memref_lowering}
\end{minipage}
\end{figure}

Next, the \texttt{SCFToEmitCPass} resolves the control flow constraints imposed by MLIR's Static Single Assignment (SSA) form. It deconstructs the \texttt{scf.for} block into an imperative \texttt{emitc.for} loop, replacing the SSA loop-carried parameters with mutable C variables (\texttt{emitc.variable}) and direct assignment operations (\texttt{emitc.assign}).

Continuing with the lowering, the \texttt{RVVToEmitCPass} translates the vector operations into C intrinsics. To achieve this, our design defines a custom \texttt{rvv} MLIR dialect within xDSL by subclassing the \texttt{IRDLOperation} class for each hardware instruction, see Figure \ref{lst:irdl_def}. For instance, the FMA instruction \texttt{vfmacc\_vf\_f32m1Op} (line 2) is constructed by declaring its required inputs using \texttt{operand\_def} (lines 5--9) and its output using \texttt{result\_def} (line 10). 

\begin{figure}[tb]
\centering
\begin{minipage}[t]{\textwidth}
\begin{lstlisting}[language=python, basicstyle=\scriptsize\ttfamily, numbers=left, frame=lines]
@irdl_op_definition
class vfmacc_vf_f32m1Op(IRDLOperation):
    name = "rvv.vfmacc_vf_f32m1Op"
    # Statically define and enforce structural data type constraints
    vd     = operand_def(RVVFloat32M1Type)    # Accumulator vector register
    memref = operand_def(MemRefType)          # Abstract multidimensional memory pointer
    offset = operand_def(IndexType)           # Integer offset to extract scalar
    vs     = operand_def(RVVFloat32M1Type)    # Fetched vector register
    avl    = operand_def(IndexType)           # Integer vector length bound
    result = result_def(RVVFloat32M1Type)     # Returning accumulator
\end{lstlisting}
\caption{xDSL definition of the \texttt{vfmacc\_vf\_f32m1Op} dialect operation.}
\label{lst:irdl_def}
\end{minipage}
\end{figure}

During this lowering stage, the \texttt{RVVToEmitCPass} utilizes the \texttt{RewritePattern} infrastructure to map these constructs into \texttt{emitc} nodes; see Figure \ref{lst:rewriter}. At its core is the \texttt{match\_and\_rewrite} function (lines 3--20), which intercepts instances of the custom \texttt{rvv} operations during tree traversal. When the rewriter encounters a \texttt{vfmacc\_vf\_f32m1Op}, it extracts the SSA operands, such as the accumulator, the memory inputs, and the vector length (lines 11--14). The rewriter then constructs the equivalent structural representation, injects \texttt{EmitCSubscriptOp} nodes to calculate memory pointers, and builds an \texttt{emitc.CallOpaqueOp} operation (lines 8--17) to bridge the functional gap. This instructs the translator to emit the raw string literal of the C hardware intrinsic: \texttt{"\_\_riscv\_vfmacc\_vf\_f32m1"} (line 9).

\begin{figure}[tb]
\centering
\begin{minipage}[t]{\textwidth}
\begin{lstlisting}[language=python, basicstyle=\scriptsize\ttfamily, numbers=left, frame=lines]
class ConvertRVV_vfmacc_vf_f32m1_ToEmitC(RewritePattern):
    @op_type_rewrite_pattern
    def match_and_rewrite(self, op: vfmacc_vf_f32m1Op, rewriter: PatternRewriter):
        # 1. Define the resulting C struct type as an opaque wrapper
        vector_type = emitc.EmitC_OpaqueType(StringAttr("vfloat32m1_t"))
        
        # 2. Construct the literal C intrinsic call with the intercepted MLIR SSA operands
        call_op = emitc.EmitC_CallOpaqueOp(
            callee="__riscv_vfmacc_vf_f32m1",
            call_args=[
                op.vd,           # Accumulator Cr vector
                load_op.result,  # Loaded Bc scalar (omitted subscript logic for brevity)
                op.vs,           # Loaded Ar vector
                op.avl           # Vector length
            ],
            result_types=[vector_type],
        )
        
        # 3. Replace the abstract custom dialect operation with the concrete EmitC node
        rewriter.replace_op(op, call_op)
\end{lstlisting}
\caption{Rewrite pattern to lower FMA instructions to EmitC opaque calls for RVV intrinsics.}
\label{lst:rewriter}
\end{minipage}
\end{figure}


\begin{figure}[tb]
\centering
\begin{minipage}[t]{\textwidth}
\begin{lstlisting}[language=C, basicstyle=\scriptsize\ttfamily, numbers=left, frame=lines]
// Lowered RVV FMA intrinsic preserving the opaque target data types
%74 = "emitc.call_opaque"(%71, %73, %62, %59) <{callee = "__riscv_vfmacc_vf_f32m1"}> 
      : (!emitc.opaque<"vfloat32m1_t">, f32, !emitc.opaque<"vfloat32m1_t">, index) 
      -> !emitc.opaque<"vfloat32m1_t">
\end{lstlisting}
\caption{\texttt{Emitc} representation of the lowered FMA instruction.}
\label{lst:emitc_opaque}
\end{minipage}
\end{figure}

The fully lowered FMA operation appears inside the \texttt{emitc} code as shown in Figure \ref{lst:emitc_opaque}. The \texttt{emitc} opaque calls encapsulate both the translated C strings and the data type constraints. We can deconstruct the four concretized arguments passed into this emitted hardware call:
\texttt{\%71 (!emitc.opaque<"vfloat32m1\_t">)} is the accumulator $C_r$ register and  the SSA pass-through (\texttt{\%42}) is transformed into a loaded mutable state variable (\texttt{\%71}); \texttt{\%73 (f32)} is the scalar extracted from the $B_c$ vector; 
\texttt{\%62 (!emitc.opaque<"vfloat32m1\_t">)} is the $A_r$ register, preserving the opaque hardware struct type; and \texttt{\%59 (index)} is the vector length controlling the operation.

~\\ \noindent
\textbf{Stage 4: C++ Translation.}
With the IR fully flattened into \texttt{emitc} nodes, we exit the xDSL environment and invoke the official MLIR \texttt{mlir-translate} tool. This tool walks the \texttt{emitc}  syntax and generates the final C code. 

The translation of loop L6 is in Figure~\ref{lst:cpp_output}, where the complex SSA yield mechanisms are lowered into standard mutable variable updates, and the FMA math operations are represented as native RISC-V hardware intrinsics.

\begin{figure}[tb]
\centering
\begin{minipage}[t]{\textwidth}
\begin{lstlisting}[language=C, basicstyle=\scriptsize\ttfamily, numbers=left, frame=lines]
// Mutable stateful accumulators replacing SSA mechanisms (using lowered C types)
  vfloat32m1_t v42 = v21;
  ...
  // Standard Imperative  kc-Loop
  for (size_t i50 = v11; i50 < v1; i50 += v12) {                    // L6 loop
    // RVV Load Intrinsic
    vfloat32m1_t v54 = __riscv_vle32_v_f32m1(v53, v52);             // Ac loading
    
    // RVV FMA Intrinsic (Computing Col 0)
    vfloat32m1_t v63 = v42;    
    float v64 = v5[v55];       
    vfloat32m1_t v65 = __riscv_vfmacc_vf_f32m1(v63, v64, v54, v52); // FMA
    
    // Overwriting the mutable variables for the next loop iteration
    v42 = v65;
    ...
  }
\end{lstlisting}
\caption{Final emitted C code implementing the hardware-accelerated loop L6.}
\label{lst:cpp_output}
\end{minipage}
\end{figure}

We can visibly trace how the final compilation stage maps the opaque \texttt{emitc} data types strictly as native C \texttt{vfloat32m1\_t} structs. Furthermore, the five original arguments from our abstract MLIR \texttt{rvv.vfmacc\_vf\_f32m1Op} have seamlessly resolved into the exactly four concrete variables required by the \texttt{vfmacc} hardware intrinsic: \texttt{v63 (vfloat32m1\_t)} is the running $C_r$ accumulator vector, loaded directly from the mutable loop proxy variable (\texttt{v42}); \texttt{v64 (float)} is the explicitly extracted scalar $B_c$ float value. Note how \texttt{emitc} successfully resolved the dual \texttt{memref}/\texttt{offset} MLIR arguments by mapping them directly to a standard C hardware array (\texttt{v5[v55]}); \texttt{v54 (vfloat32m1\_t)} is the loaded $Ar$ vector register; and \texttt{v52 (size\_t)} is the vector length of the operation.

~\\ \noindent
\textbf{Stages 5, 6: Test Bench Generation and Deployment.}
Our pipeline automatically generates a comprehensive C test bench harness, complete with verification routines and Makefile infrastructure. It then orchestrates the deployment, native compilation, and benchmarking of the generated kernels on the target remote RISC-V hardware, providing end-to-end analysis. 

The artifact for generating micro-kernels and reproducing the results of this paper is available at \url{https://github.com/JieGH/RVV_code_gen_via_MLIR_xDSL}.
\section{Performance Evaluation}
\label{sec:experiments}

In this section, we first evaluate the performance of the generated micro-kernels in a stand-alone setting. 
Then, we assess the impact of the automatically generated micro-kernels within the complete \gemm algorithm and compare the resulting performance with the OpenBLAS library across several scenarios. The experiments presented in this work were conducted on the following platforms:

\medskip
\noindent
\textbf{K230.}
This platform corresponds to a CanMV-K230 board integrating the K230 system-on-chip (SoC). The processor follows a {\em big-LITTLE} configuration; however, only the RVV-capable C908 core was used in our experiments. This core operates at 1.6~GHz and supports the RISC-V Vector Extension (RVV) version 1.0 with a 128-bit vector unit. The processor includes 32~KB of L1 data cache and 256~KB of L2 cache.

\noindent
\textbf{BPI.}
The BananaPi F3 board integrates an 8-core SpaceMiT K1 processor operating at 1.6~GHz. Each core features a 32~KB L1 data cache, while a 512~KB L2 cache is shared among clusters of four cores. The processor implements the RVV 1.0 specification with a vector length of 256 bits.

Experiments used Python \texttt{3.13.9}, xDSL \texttt{v0.54.3}, and LLVM \texttt{v22.0}. Generated C code was compiled with \texttt{g++-14 -march=rv64gcvzfh -mabi=lp64d}. We compared against OpenBLAS \texttt{v0.3.31}, compiled directly for the K230 \\(\texttt{TARGET=RISCV64\_ZVL128B}, $8\times 8$ kernel) and BPI (\texttt{TARGET=RISCV64\_ZVL256B}, $16\times 8$ kernel). We report single-core, single-precision performance in GFLOPS, averaging 1500 runs for micro-kernels and 200 for \gemm. Parallel evaluation was excluded to isolate core architectural effects.



\subsection{Micro-kernel evaluation}

\begin{figure}[tb]
    \centering
    \includegraphics[width=0.47\textwidth]{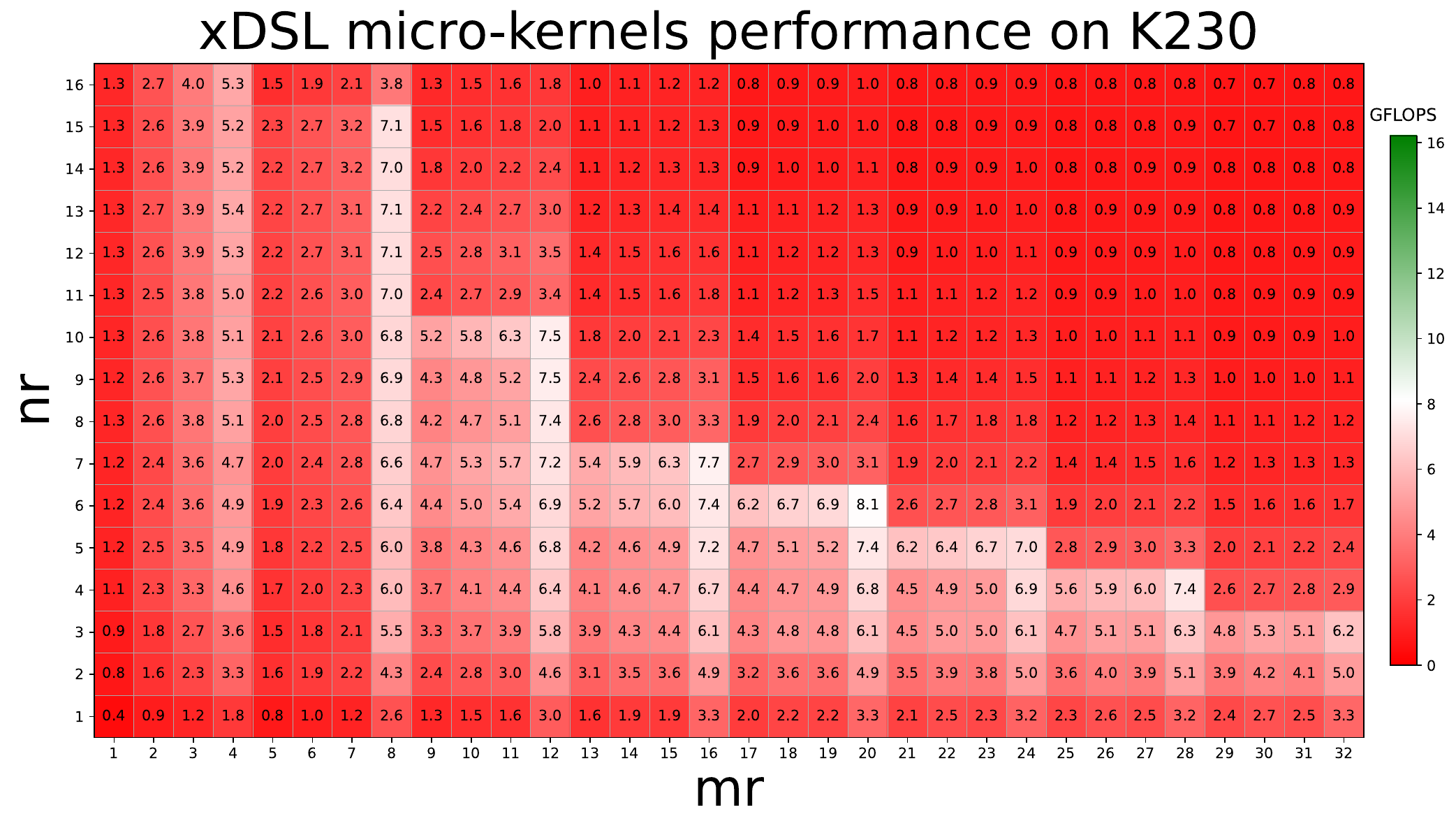}
    \includegraphics[width=0.47\textwidth]{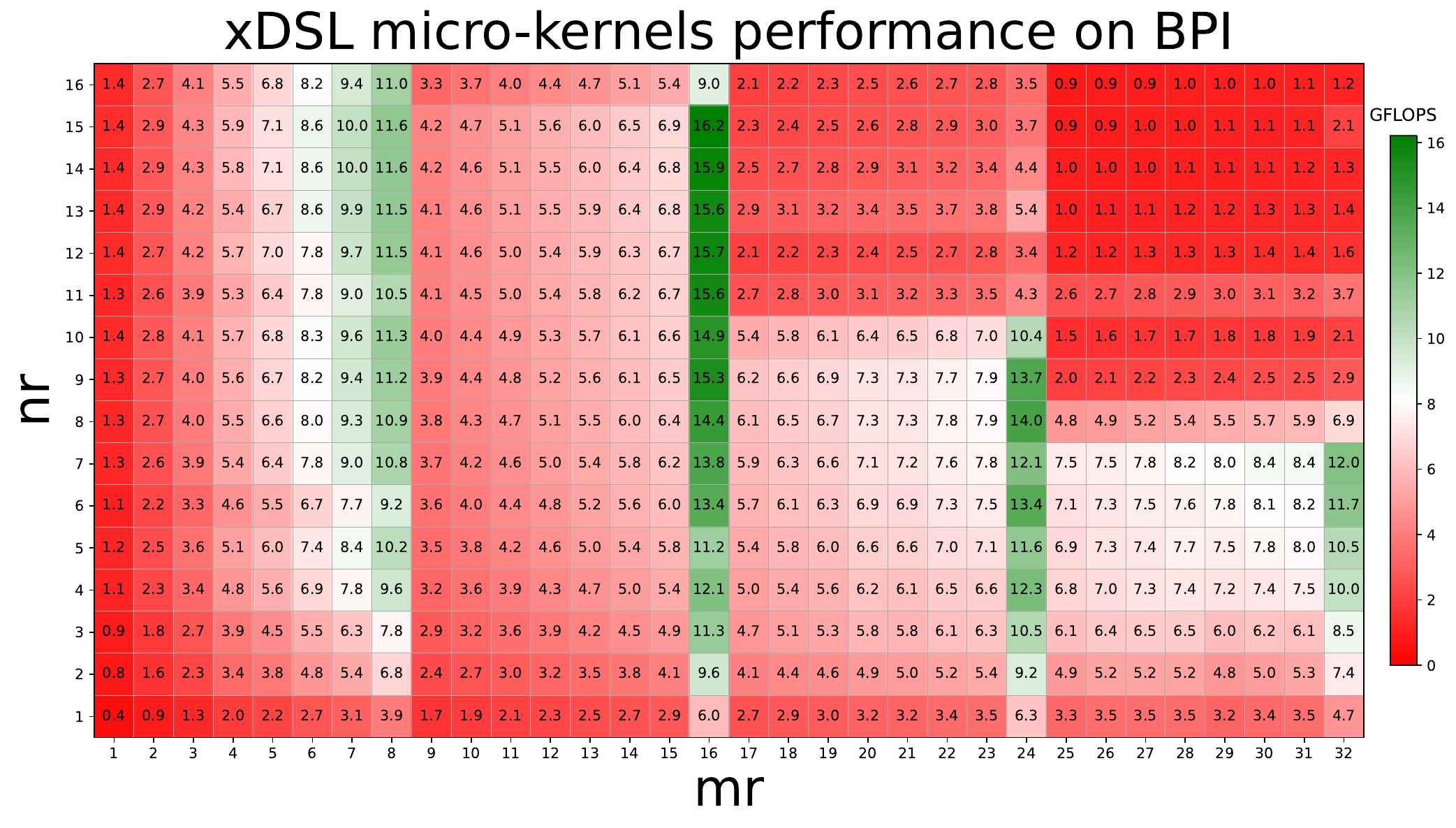}
    \caption{Micro-kernels performance for K230 (left) and BPI (right).}
\label{figures:analyze_speedup_xdsl_perf_heatmap}
\end{figure}

Figure~\ref{figures:analyze_speedup_xdsl_perf_heatmap} summarizes the performance of different micro-kernel configurations evaluated in isolation.  The results reveal a clear relationship between micro-kernel dimensions and vector utilization. In particular, configurations in which the $m_r$ dimension is a multiple of the number of elements that fit into a vector register achieve higher performance. For FP32 operands, this corresponds to multiples of 4 elements on K230 and multiples of 8 elements on BPI. 

Among the evaluated configurations, the best micro-kernel dimensions are $20 \times 6$ for K230 and $16 \times 15$ for BPI, achieving peak performances of 8.1~GFLOPS and 16.2~GFLOPS, respectively. Notice that the range of values is up to 16.2~GFLOPS for both charts, revealing the performance difference between devices.

\subsection{\gemm evaluation}

To evaluate the impact of micro-kernels on the \gemm implementation, we first analyze performance using square matrices, which constitute a standard benchmark configuration widely used to evaluate dense linear algebra kernels. Second, we evaluate the algorithm using the \gemm shapes that arise in inference for the BERT-Large model. This dual evaluation allows us to assess both the peak performance of the implementation and its effectiveness in realistic machine learning. Table~\ref{tab:layer_dims} highlights the different \gemm scenarios for the evaluation. 

We adopted the established \gemm implementation on \url{https://github.com/martineh/gemm_blis_family}. This design provided a cache-efficient \gemm implementation that allows us to directly plug in our xDSL-generated kernel code and evaluate its performance. 

\setlength{\tabcolsep}{5pt}
\begin{table}[tb]
\centering
\begin{tabular}{llrrr|llrrr}
\toprule
\textbf{Case} & \textbf{ID} & \textbf{m} & \textbf{n} & \textbf{k} &
\textbf{Case} & \textbf{ID} & \textbf{m} & \textbf{n} & \textbf{k}\\
\midrule
 & S1 & 1000 & 1000 & 1000 &  & B1 & 1024 & 384 & 1024 \\
 & S2 & 2000 & 2000 & 2000 & & B2 & ~384 & 384 & ~~64  \\
 Square & S3 & 3000 & 3000 & 3000 & BERT & B3 & ~~64 & 384 & ~384  \\
 & S4 & 4000 & 4000 & 4000 & & B4 & 4096 & 384 & 1024  \\
 & S5 & 5000 & 5000 & 5000 & & B5 & 1024 & 384 & 4096  \\
\bottomrule
\end{tabular}%
\caption{Layer dimensions ($m, n, k$) for the evaluated scenarios.}
\label{tab:layer_dims}
\end{table}

\begin{figure*}[tb]
\centering
\includegraphics[width=0.49\textwidth]{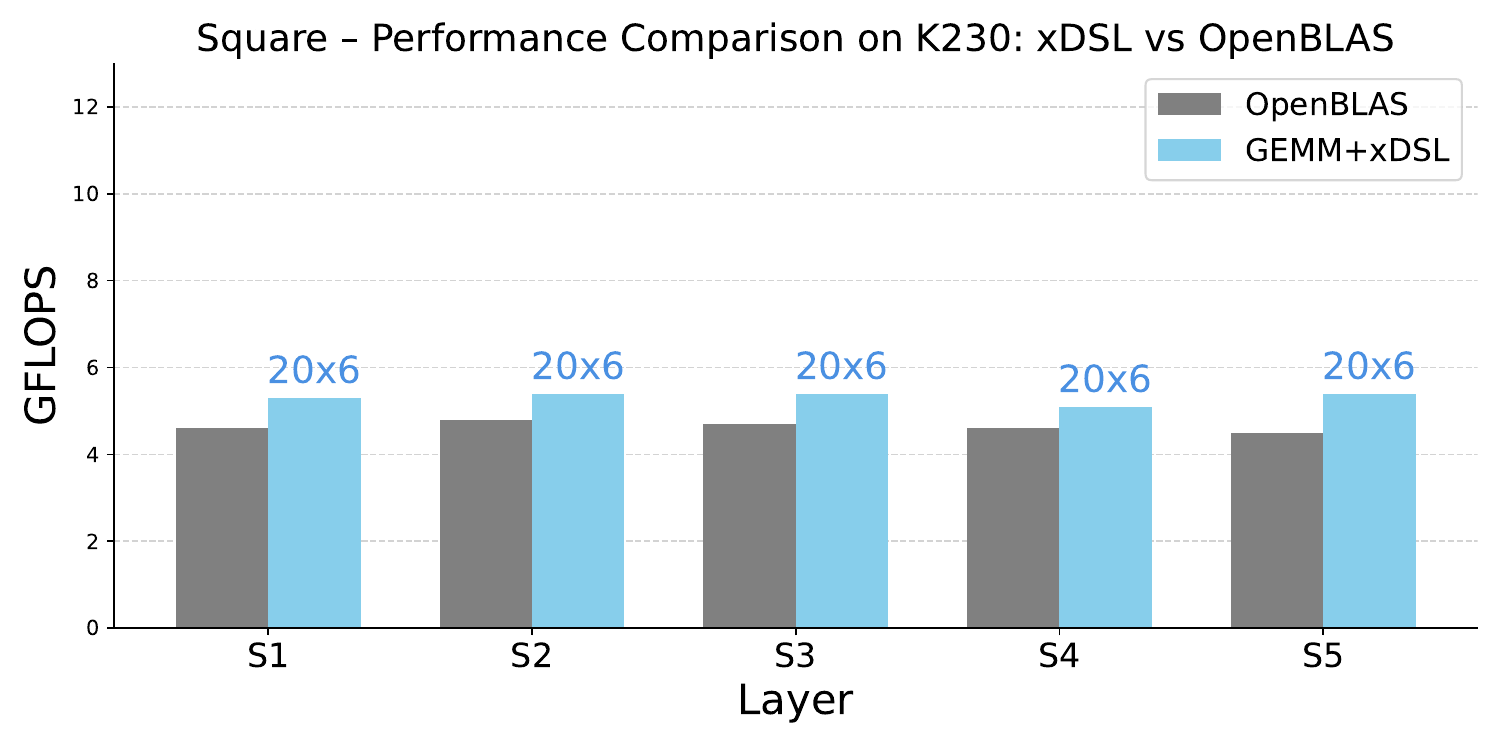} 
\includegraphics[width=0.49\textwidth]{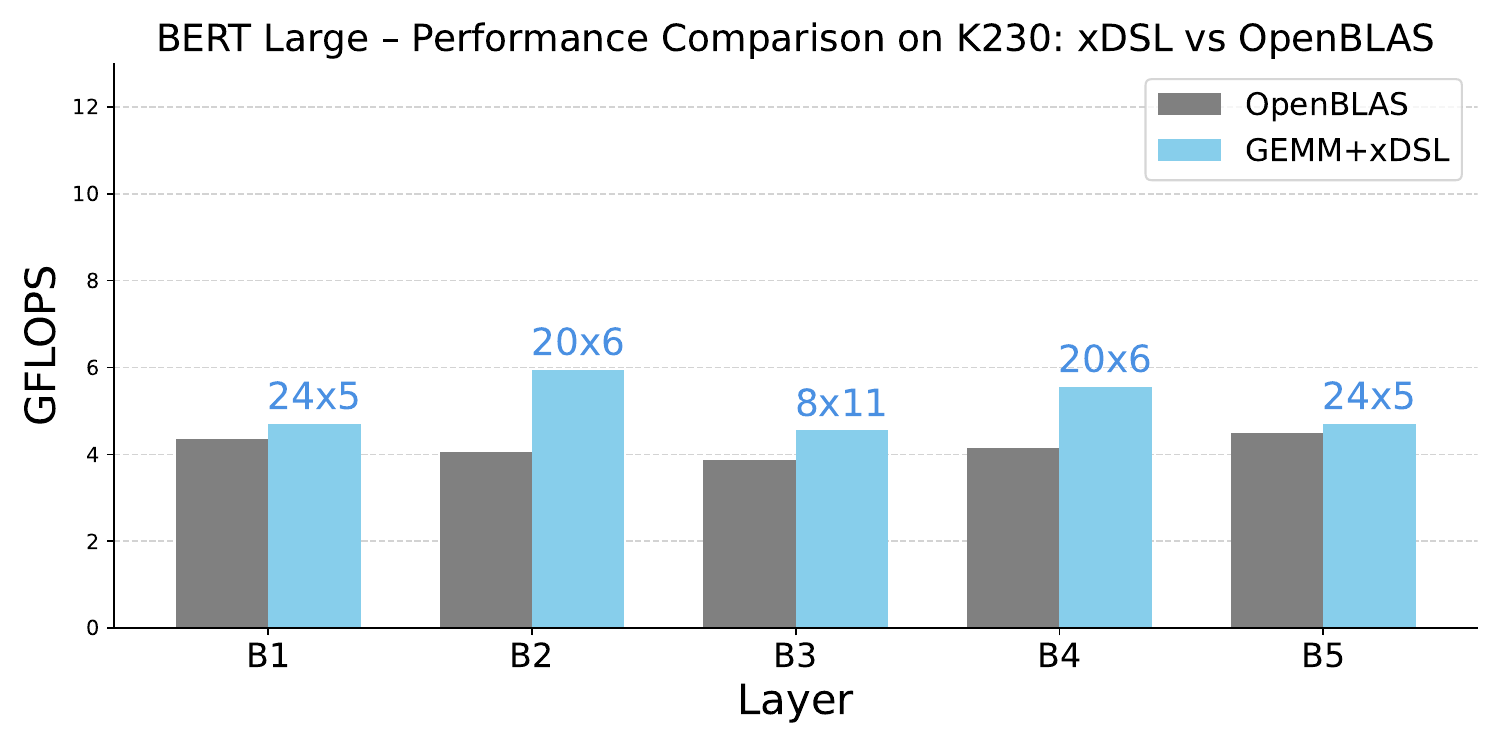} \\
\includegraphics[width=0.49\textwidth]{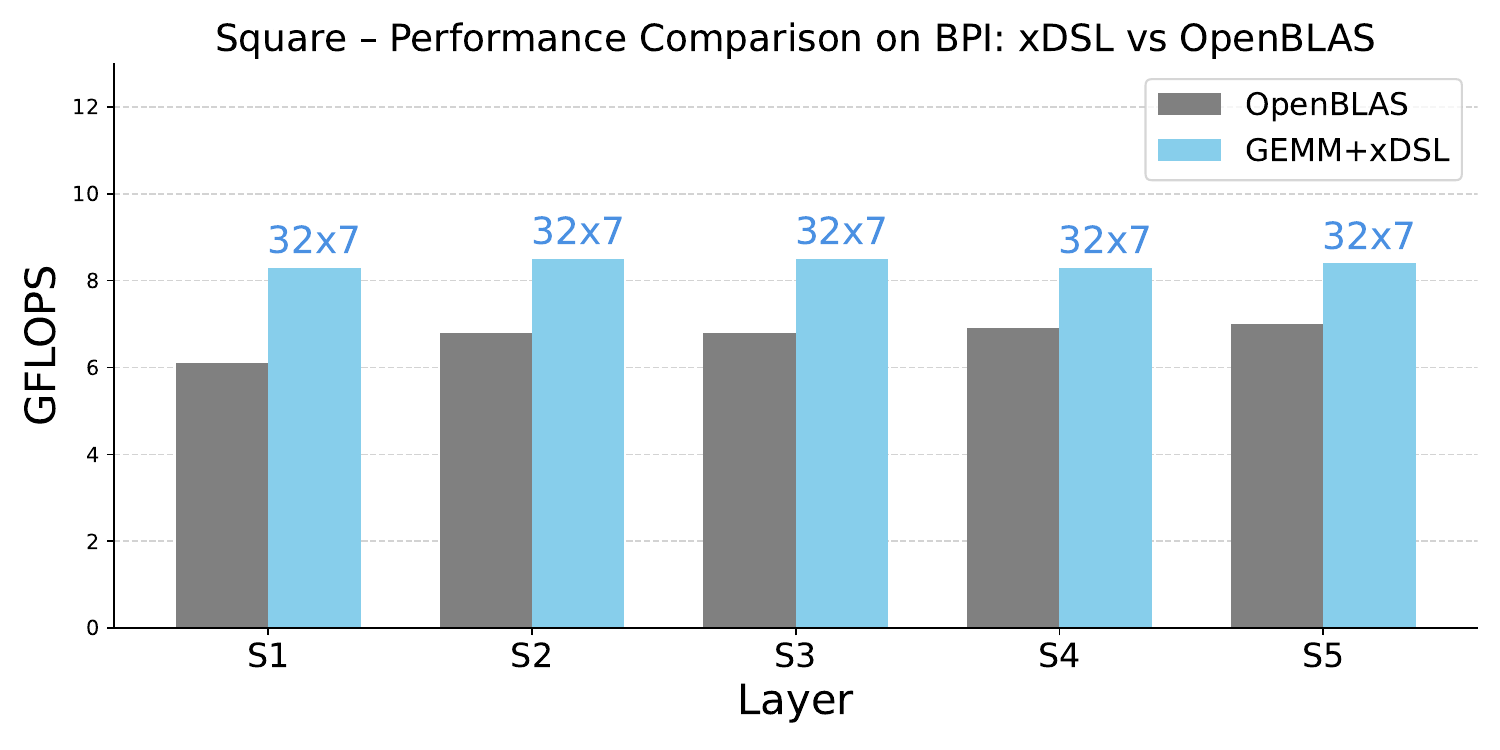} 
\includegraphics[width=0.49\textwidth]{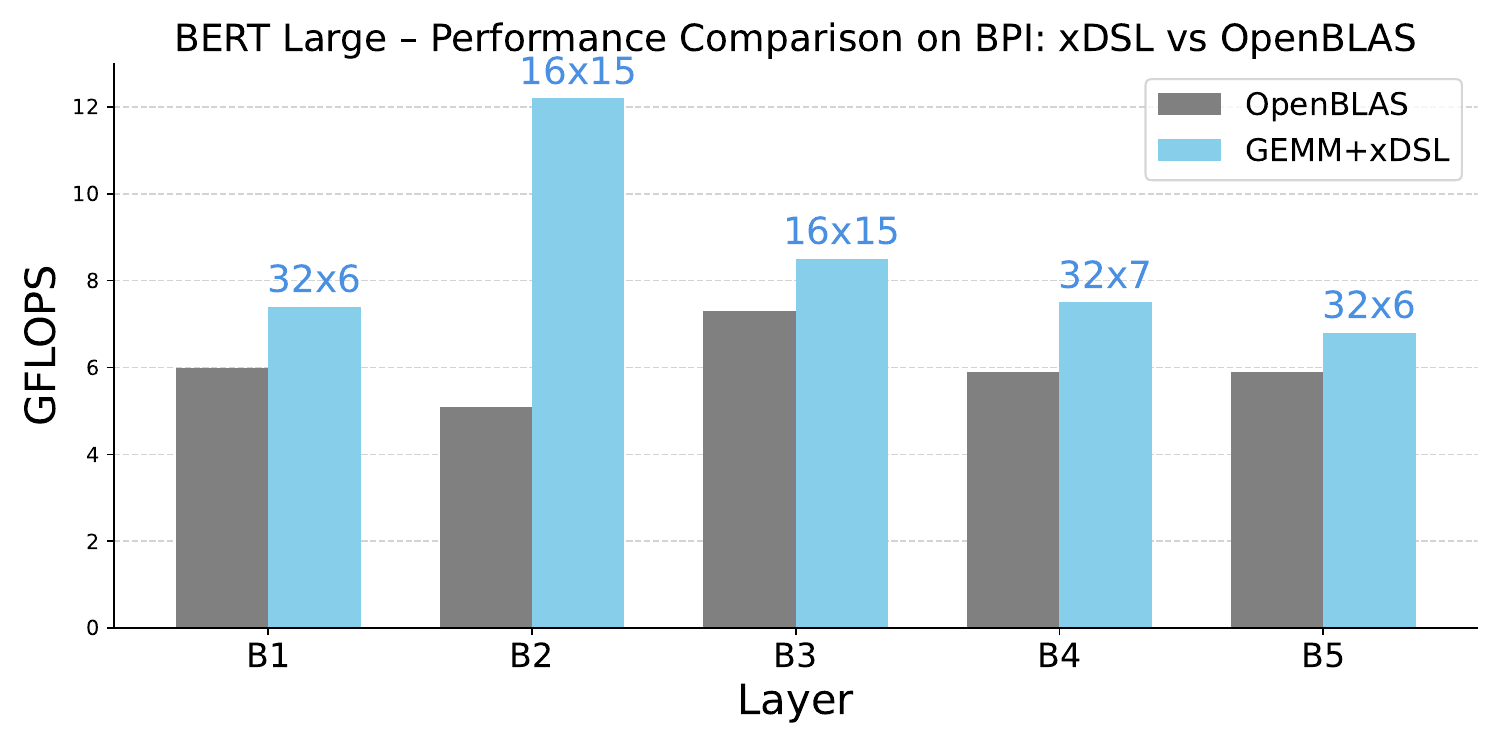} 
\caption{Comparison between \gemm with xDSL micro-kernels against OpenBLAS on K230 (top) and on BPI (bottom), for square matrices (left) and BERT (right).}
\label{fig:perf}
\end{figure*}

Figure~\ref{fig:perf} presents the performance comparison between the proposed xDSL implementation and OpenBLAS across the two workloads. 

For the square workload on K230, GEMM+xDSL consistently outperforms OpenBLAS across all evaluated layers (S1--S5). OpenBLAS achieves performance values between approximately 4.5 and 4.9 GFLOPS, whereas xDSL reaches between 5.1 and 5.4 GFLOPS. This corresponds to an average improvement of roughly 10--15\%. Notably, the optimized implementation uses the same tiling configuration (20$\times$6) for all layers, indicating that this configuration provides stable and efficient utilization of the hardware resources for matrices with regular shapes.
For the BERT Large workload on the same platform, the performance improvements are more variable but remain favorable to the proposed approach. OpenBLAS achieves between 3.9 and 4.5 GFLOPS, while xDSL reaches between 4.6 and 5.9 GFLOPS depending on the layer. The largest improvement occurs in layer B2, where performance increases from 4.0 GFLOPS to 5.9 GFLOPS, corresponding to a speedup close to 47\%. Other layers exhibit improvements between 10\% and 35\%. Unlike the Square workload, different tiling configurations (e.g., 24$\times$5, 20$\times$6, and 8$\times$11) are selected for each layer, reflecting the greater diversity of matrix shapes in transformer-based models.

On BPI, the advantages of xDSL become more pronounced. For the square workload, OpenBLAS achieves between 6.1 and 7.0 GFLOPS, whereas xDSL reaches approximately 8.3--8.6 GFLOPS. This corresponds to performance improvements ranging from about 20\% to 35\%. As in the K230, a single tiling configuration (32$\times$7) is used for all layers, demonstrating that a well-chosen configuration can deliver stable performance when matrix dimensions remain relatively uniform.
The largest performance gains are observed for the BERT Large workload on BPI. OpenBLAS achieves between 5.1 and 7.3 GFLOPS, while GEMM+xDSL reaches between 6.8 and 12.2 GFLOPS. In particular, layer B2 shows a significant improvement from approximately 5.1 GFLOPS to 12.2 GFLOPS, representing more than a twofold speedup. The remaining layers show improvements ranging from roughly 15\% to 30\%. Similar to the K230 experiments, different tiling configurations (e.g., 32$\times$6, 16$\times$15, and 32$\times$7) are selected to better match the computational characteristics of each layer.

Overall, the results demonstrate that xDSL consistently outperforms OpenBLAS across both hardware platforms and workloads. The improvements are particularly significant for the transformer, where layer-specific matrix dimensions benefit from adaptive tiling strategies. Furthermore, the results suggest that the proposed approach is capable of effectively exploiting the underlying hardware characteristics, especially on the BPI platform, where performance gains exceed 2$\times$ in certain layers.

\section{Conclusions}
\label{sec:conclusions}

This paper presented a hybrid MLIR--xDSL compilation pipeline that enables the generation of optimized RISC-V Vector code starting from high-level MLIR abstractions. The proposed approach addresses a practical limitation of the current MLIR infrastructure, namely the lack of complete lowering paths that translate high-level dialects into hardware-aware RVV implementations suitable for production environments.
By leveraging the flexibility of xDSL, we implemented custom lowering passes that bridge the gap between high-level MLIR representations and the \texttt{emitc} dialect, ultimately enabling the automatic generation of portable C code invoking RVV intrinsics. 

The evaluation demonstrated that the automatically generated micro-kernels achieve performance comparable to hand-written implementations, confirming the effectiveness of the proposed code generation strategy. When integrated into a complete \gemm implementation, the generated kernels consistently outperformed the OpenBLAS baseline across both evaluated workloads. For square matrices, the proposed implementation achieved performance improvements of up to 35\%, while for transformer-based workloads derived from the BERT-Large model, the speedups reached up to 2.4$\times$. These results highlight the benefits of automatically generating multiple micro-kernel configurations and selecting the most suitable one for each workload or layer dimension.


Future work will explore extending the framework to additional linear-algebra kernels,  mixed-precision arithmetic, and the automatic generation of packing routines. 

~\\
\noindent
\textbf{Acknowledgments.} 
This work received funding from projects
PID2023-146569NB-C2 of MCIN/AEI/10.13 039/501100011033,
and CIPROM/2022/20 of the Generalitat Valenciana, 
the DARE SGA1 Project, from the European High-Performance Computing Joint Undertaking (JU) under Grant Agreement No 101202459 and from PCI2024-161687-3 Project funded by MICIU/AEI/10.13039/ 501100011033 and European's Union “NextGenerationEU”/PRTR”. The JU receives support from the European Union’s Horizon Europe research and innovation programme and Spain, Germany, Czechia, Italy, Netherlands, Belgium, Finland, Greece, Croatia, Portugal, Poland, Sweden, France and Austria.
\bibliographystyle{splncs04}
\bibliography{Bibliography}

@inproceedings{siwinska2025enhancing,
  title={Enhancing Transformer Performance and Portability through Auto-tuning Frameworks},
  author={Siwinska, P and Lei, J and Castell{\'o}, A and Alonso-Jord{\'a}, P and Quintana-Ort{\'\i}, ES},
  booktitle={The Journal of Supercomputing},
   year={2025},
isbn = {1573-0484},
OPTurl = {https://doi.org/10.1007/s11227-026-08327-6},
doi = {10.1007/s11227-026-08327-6},
}

@Misc{RVV,
  key =          {{RISC-V} V Vector extension},
  title =          {{RISC-V} V vector extension},
  OPTauthor =    {},
  OPTtitle =     {},
  howpublished = {\url{https://github.com/riscv/riscv-v-spec/releases/download/v1.0/riscv-v-spec-1.0.pdf}},
  OPTmonth =     {},
  OPTyear =         {},
  OPTnote =      {},
  OPTannote =    {}
}

@inproceedings{Halide,
author = {Ragan-Kelley, Jonathan and Barnes, Connelly and Adams, Andrew and Paris, Sylvain and Durand, Fr\'{e}do and Amarasinghe, Saman},
title = {Halide: a language and compiler for optimizing parallelism, locality, and recomputation in image processing pipelines},
year = {2013},
isbn = {9781450320146},
publisher = {Association for Computing Machinery},
address = {New York, NY, USA},
url = {https://doi.org/10.1145/2491956.2462176},
doi = {10.1145/2491956.2462176},
booktitle = {Proceedings of the 34th ACM SIGPLAN Conference on Programming Language Design and Implementation},
pages = {519–530},
numpages = {12},
location = {Seattle, Washington, USA},
series = {PLDI '13}
}

@INPROCEEDINGS{mlir2,
  author={Lattner, Chris and Amini, Mehdi and Bondhugula, Uday and Cohen, Albert and Davis, Andy and Pienaar, Jacques and Riddle, River and Shpeisman, Tatiana and Vasilache, Nicolas and Zinenko, Oleksandr},
  booktitle={2021 IEEE/ACM International Symposium on Code Generation and Optimization (CGO)}, 
  title={MLIR: Scaling Compiler Infrastructure for Domain Specific Computation}, 
  year={2021},
  volume={},
  number={},
  pages={2-14},
  doi={10.1109/CGO51591.2021.9370308}}

@Misc{OpenBLAS,
  key =          {{O}pen{BLAS}},
  title =          {{O}pen{BLAS}},
  OPTauthor =    {},
  OPTtitle =     {},
  OPTmonth =     {},
  year =         {2012},
  OPTnote =      {},
  howpublished = {\url{http://xianyi.github.com/OpenBLAS/}},
  OPTannote =    {}
}

@inproceedings{10.1145/3696443.3708952,
author = {Lopoukhine, Alexandre and Ficarelli, Federico and Vasiladiotis, Christos and Lydike, Anton and Van Delm, Josse and Dutilleul, Alban and Benini, Luca and Verhelst, Marian and Grosser, Tobias},
title = {A Multi-level Compiler Backend for Accelerated Micro-kernels Targeting {RISC-V ISA} Extensions},
year = {2025},
isbn = {9798400712753},
publisher = {Association for Computing Machinery},
OPTaddress = {New York, NY, USA},
OPTurl = {https://doi.org/10.1145/3696443.3708952},
doi = {10.1145/3696443.3708952},
booktitle = {ACM/IEEE Int. Symp. Code Gen. \& Optimization},
pages = {163–178},
numpages = {16},
location = {Las Vegas, NV, USA},
series = {CGO '25}
}

@article{MLIR,
  author    = {Chris Lattner and
               Jacques A. Pienaar and
               Mehdi Amini and
               Uday Bondhugula and
               River Riddle and
               Albert Cohen and
               Tatiana Shpeisman and
               Andy Davis and
               Nicolas Vasilache and
               Oleksandr Zinenko},
  title     = {{MLIR:} {A} Compiler Infrastructure for the End of {M}oore's Law},
  journal   = {CoRR},
  OPTvolume    = {abs/2002.11054},
  year      = {2020},
  url       = {https://arxiv.org/abs/2002.11054},
  eprinttype = {arXiv},
  eprint    = {2002.11054},
  bibsource = {dblp computer science bibliography, https://dblp.org}
}

@article{GEMM_MLIR,
  author    = {Uday Bondhugula},
  title     = {High Performance Code Generation in {MLIR:} An Early Case Study with
               {GEMM}},
  journal   = {CoRR},
  volume    = {abs/2003.00532},
  year      = {2020},
  url       = {https://arxiv.org/abs/2003.00532},
  eprinttype = {arXiv},
  eprint    = {2003.00532},
  bibsource = {dblp computer science bibliography, https://dblp.org}
}

@inproceedings{yukaexo,
author = {Ikarashi, Yuka and Bernstein, Gilbert Louis and Reinking, Alex and Genc, Hasan and Ragan-Kelley, Jonathan},
title = {Exocompilation for Productive Programming of Hardware Accelerators},
year = {2022},
isbn = {9781450392655},
publisher = {Association for Computing Machinery},
OPTaddress = {New York, NY, USA},
OPTurl = {https://doi.org/10.1145/3519939.3523446},
doi = {10.1145/3519939.3523446},
booktitle = {Proceedings of the 43rd ACM SIGPLAN International Conference on Programming Language Design and Implementation},
pages = {703–718},
numpages = {16},
location = {San Diego, CA, USA},
series = {PLDI}
}

@article{Goto:2008:AHP,
 author = {Goto, Kazushige and Geijn, Robert A. van de},
 title = {Anatomy of High-performance Matrix Multiplication},
 journal = {ACM Trans. Math. Softw.},
 issue_date = {May 2008},
 volume = {34},
 number = {3},
 month = may,
 year = {2008},
 issn = {0098-3500},
 pages = {12:1--12:25},
 articleno = {12},
 numpages = {25},
 OPTurl = {http://doi.acm.org/10.1145/1356052.1356053},
 doi = {10.1145/1356052.1356053},
 acmid = {1356053},
 publisher = {ACM},
 address = {New York, NY, USA},
}

@conference{xdsl,
title = "{xDSL}: A common compiler ecosystem for domain specific languages",
author = "Nick Brown and Tobias Grosser and Mathieu Fehr and Michel Steuwer and Paul Kelly",
year = "2022",
month = nov,
day = "1",
booktitle = {Proceedings of the SC '23 Workshops of The Int. Conference on High Performance Computing, Network, Storage, and Analysis},
language = "English",
note = "Supercomputing 2023, SC23",
OPTurl = "https://sc23.supercomputing.org/",
}

@inproceedings{xdsl2,
author = {Fehr, Mathieu and Weber, Michel and Ulmann, Christian and Lopoukhine, Alexandre and L\"{u}cke, Martin Paul and Degioanni, Th\'{e}o and Vasiladiotis, Christos and Steuwer, Michel and Grosser, Tobias},
title = {{xDSL}: Sidekick Compilation for {SSA}-Based Compilers},
year = {2025},
isbn = {9798400712753},
publisher = {Association for Computing Machinery},
OPTaddress = {New York, NY, USA},
OPTurl = {https://doi.org/10.1145/3696443.3708945},
doi = {10.1145/3696443.3708945},
booktitle = {ACM/IEEE Int. Symp. Code Gen. \& Optimization},
pages = {179–192},
numpages = {14},
OPTkeywords = {compilation frameworks, interchange formats, intermediate representations},
location = {Las Vegas, NV, USA},
series = {CGO '25}
}

@article{TVM_1,
author    = {Tianqi Chen and
               Thierry Moreau and
               Ziheng Jiang and
               Haichen Shen and
               Eddie Q. Yan and
               Leyuan Wang and
               Yuwei Hu and
               Luis Ceze and
               Carlos Guestrin and
               Arvind Krishnamurthy},
  title     = {{TVM:} End-to-End Optimization Stack for Deep Learning},
  journal   = {CoRR},
  volume    = {abs/1802.04799},
  year      = {2018},
  url       = {http://arxiv.org/abs/1802.04799},
  eprinttype = {arXiv},
  eprint    = {1802.04799},
  bibsource = {dblp computer science bibliography, https://dblp.org}
}

@INPROCEEDINGS{10444883,
  author={Castelló, Adrián and Bellavita, Julian and Dinh, Grace and Ikarashi, Yuka and Martínez, Héctor},
  booktitle={IEEE/ACM Int. Symp. Code Gen. \& Optimization}, 
  title={Tackling the Matrix Multiplication Micro-Kernel Generation with {Exo}}, 
  year={2024},
  volume={},
  number={},
  pages={182-193},
  doi={10.1109/CGO57630.2024.10444883}}

@InProceedings{castelloISCcots,
author="Castell{\'o}, Adri{\'a}n
and Mart{\'i}nez, H{\'e}ctor
and Catal{\'a}n, Sandra
and Igual, Francisco D.
and Quintana-Ort{\'i}, Enrique S.",
OPTeditor="Neuwirth, Sarah
and Paul, Arnab Kumar
and Weinzierl, Tobias
and Carson, Erin Claire",
title="Evaluation of {RVV}-Enabled {COTS} Platforms with Matrix Multiplication and {Exo}",
booktitle="High Performance Computing. Lecture Notes in Computer Science, vol. 16091",
year="2026",
publisher="Springer Nature Switzerland",
address="Cham",
pages="534--547",
isbn="978-3-032-07612-0"
}

@article{alaejos2022micro,
  title={Micro-kernels for portable and efficient matrix multiplication in deep learning},
  author={Alaejos, Guillermo and Castell{\'o}, Adri{\'a}n and Mart{\'\i}nez, H{\'e}ctor and Alonso-Jord{\'a}, Pedro and Igual, Francisco D and Quintana-Ort{\'\i}, Enrique S},
  journal={The Journal of Supercomputing},
  pages={1--24},
  year={2022},
  publisher={Springer}
}

\end{document}